\documentclass[tighten,twocolumn]{aastex7}

\begin{document}

\title{Extremely Metal-Poor Galaxies in DESI DR1: Connections to Galaxies in the Early Universe}

\correspondingauthor{Hu Zou}

\author[orcid=0000-0003-2132-0727,sname='Sui']{Jipeng Sui}
\affiliation{National Astronomical Observatories, Chinese Academy of Sciences, Beijing 100101, China}
\affiliation{School of Astronomy and Space Science, University of Chinese Academy of Sciences, Beijing 101408, China}
\email{suijp@bao.ac.cn}  

\author[orcid=0000-0002-6684-3997,gname=Hu, sname='Zou']{Hu Zou} 
\affiliation{National Astronomical Observatories, Chinese Academy of Sciences, Beijing 100101, China}
\affiliation{School of Astronomy and Space Science, University of Chinese Academy of Sciences, Beijing 101408, China}
\email[show]{zouhu@nao.cas.cn}

\author{Dirk Scholte}
\affiliation{Institute for Astronomy, University of Edinburgh, Royal Observatory, Blackford Hill, Edinburgh EH9 3HJ, UK}
\email{dscholte@roe.ac.uk}

\author{Amélie Saintonge}
\affiliation{Department of Physics \& Astronomy, University College London, Gower Street, London, WC1E 6BT, UK}
\email{a.saintonge@ucl.ac.uk}

\author{Mar Mezcua}
\affiliation{Institute of Space Sciences, ICE-CSIC, Campus UAB, Carrer de Can Magrans s/n, 08913 Bellaterra, Barcelona, Spain}
\affiliation{Institut d'Estudis Espacials de Catalunya (IEEC), c/ Esteve Terradas 1, Edifici RDIT, Campus PMT-UPC, 08860 Castelldefels, Spain}
\email{mezcua@ice.csic.es}

\author{Malgorzata Siudek}
\affiliation{Instituto de Astrof\'{\i}sica de Canarias, V\'{\i}a L\'actea, 38205 La Laguna, Tenerife, Spain}
\affiliation{Instituto de Astrof\'isica de Canarias (IAC); Departamento de Astrof\'isica, Universidad de La Laguna (ULL), 38200, La Laguna, Tenerife, Spain}
\email{malgorzata.siudek@iac.es}

\author{Wenxiong Li}
\affiliation{National Astronomical Observatories, Chinese Academy of Sciences, Beijing 100101, China}
\email{}

\author{Wei-Jian Guo}
\affiliation{National Astronomical Observatories, Chinese Academy of Sciences, Beijing 100101, China}
\email{}

\author{Shufei Liu}
\affiliation{National Astronomical Observatories, Chinese Academy of Sciences, Beijing 100101, China}
\affiliation{School of Astronomy and Space Science, University of Chinese Academy of Sciences, Beijing 101408, China}
\email{}

\author{Yunao Xiao}
\affiliation{National Astronomical Observatories, Chinese Academy of Sciences, Beijing 100101, China}
\affiliation{School of Astronomy and Space Science, University of Chinese Academy of Sciences, Beijing 101408, China}
\email{}

\author{Francisco Prada}
\affiliation{Instituto de Astrof\'{i}sica de Andaluc\'{i}a (CSIC), Glorieta de la Astronom\'{i}a, s/n, E-18008 Granada, Spain}
\email{fprada@iaa.es}

\author[orcid=0000-0002-3983-6484]{Siwei Zou}
\affiliation{Chinese Academy of Sciences South America Center for Astronomy, National Astronomical Observatories, CAS, Beijing 100101, China}
\email{zousw@nao.cas.cn}

\author{Jessica Nicole Aguilar}
\affiliation{Lawrence Berkeley National Laboratory, 1 Cyclotron Road, Berkeley, CA 94720, USA}
\email{jaguilar@lbl.gov}

\author{Steven Ahlen}
\affiliation{Department of Physics, Boston University, 590 Commonwealth Avenue, Boston, MA 02215 USA}
\email{ahlen@bu.edu}

\author{Carlos Allende Prieto}
\affiliation{Instituto de Astrof\'{\i}sica de Canarias, C/ V\'{\i}a L\'{a}ctea, s/n, E-38205 La Laguna, Tenerife, Spain}
\affiliation{Departamento de Astrof\'{\i}sica, Universidad de La Laguna (ULL), E-38206, La Laguna, Tenerife, Spain}
\email{carlos.allende.prieto@iac.es}

\author{Davide Bianchi}
\affiliation{Dipartimento di Fisica ``Aldo Pontremoli'', Universit\`a degli Studi di Milano, Via Celoria 16, I-20133 Milano, Italy}
\affiliation{INAF-Osservatorio Astronomico di Brera, Via Brera 28, 20122 Milano, Italy}
\email{davide.bianchi1@unimi.it}

\author{David Brooks}
\affiliation{Department of Physics \& Astronomy, University College London, Gower Street, London, WC1E 6BT, UK}
\email{david.brooks@ucl.ac.uk}

\author{Yu-Ling Chang}
\affiliation{Graduate Institute of Astrophysics and Department of Physics, National Taiwan University, No. 1, Sec. 4, Roosevelt Rd., Taipei 10617, Taiwan, China}
\email{ecylchang@ntu.edu.tw}

\author{Todd Claybaugh}
\affiliation{Lawrence Berkeley National Laboratory, 1 Cyclotron Road, Berkeley, CA 94720, USA}
\email{tmclaybaugh@lbl.gov}

\author{Andrei Cuceu}
\affiliation{Lawrence Berkeley National Laboratory, 1 Cyclotron Road, Berkeley, CA 94720, USA}
\email{acuceu@lbl.gov}

\author{Axel de la Macorra}
\affiliation{Instituto de F\'{\i}sica, Universidad Nacional Aut\'{o}noma de M\'{e}xico, Circuito de la Investigaci\'{o}n Cient\'{\i}fica, Ciudad Universitaria, Cd. de M\'{e}xico C.~P.~04510, M\'{e}xico}
\email{macorra@fisica.unam.mx}

\author{Peter Doel}
\affiliation{Department of Physics \& Astronomy, University College London, Gower Street, London, WC1E 6BT, UK}
\email{apd@star.ucl.ac.uk}

\author{Jaime E. Forero-Romero}
\affiliation{Departamento de F\'isica, Universidad de los Andes, Cra. 1 No. 18A-10, Edificio Ip, CP 111711, Bogot\'a, Colombia}
\affiliation{Observatorio Astron\'omico, Universidad de los Andes, Cra. 1 No. 18A-10, Edificio H, CP 111711 Bogot\'a, Colombia}
\email{je.forero@uniandes.edu.co}

\author{Enrique Gaztañaga}
\affiliation{Institute of Cosmology and Gravitation, University of Portsmouth, Dennis Sciama Building, Portsmouth, PO1 3FX, UK}
\affiliation{Institut d'Estudis Espacials de Catalunya (IEEC), c/ Esteve Terradas 1, Edifici RDIT, Campus PMT-UPC, 08860 Castelldefels, Spain}
\email{gaztanaga@gmail.com}

\author{Satya Gontcho A Gontcho}
\affiliation{Lawrence Berkeley National Laboratory, 1 Cyclotron Road, Berkeley, CA 94720, USA}
\affiliation{University of Virginia, Department of Astronomy, Charlottesville, VA 22904, USA}
\email{satya@virginia.edu}

\author{Gaston Gutierrez}
\affiliation{Fermi National Accelerator Laboratory, PO Box 500, Batavia, IL 60510, USA}
\email{gaston@fnal.gov}

\author{ChangHoon Hahn}
\affiliation{University of Texas at Austin, Austin, TX 78712, USA}
\email{changhoon.hahn@utexas.edu}

\author{Dick Joyce}
\affiliation{NSF NOIRLab, 950 N. Cherry Ave., Tucson, AZ 85719, USA}
\email{richard.joyce@noirlab.edu}

\author{Robert Kehoe}
\affiliation{Department of Physics, Southern Methodist University, 3215 Daniel Avenue, Dallas, TX 75275, USA}
\email{kehoe@physics.smu.edu}

\author{Martin Landriau}
\affiliation{Lawrence Berkeley National Laboratory, 1 Cyclotron Road, Berkeley, CA 94720, USA}
\email{mlandriau@lbl.gov}

\author{Laurent Le Guillou}
\affiliation{Sorbonne Universit\'{e}, CNRS/IN2P3, Laboratoire de Physique Nucl\'{e}aire et de Hautes Energies (LPNHE), FR-75005 Paris, France}
\email{llg@lpnhe.in2p3.fr}

\author{Yifei Luo}
\affiliation{Lawrence Berkeley National Laboratory, 1 Cyclotron Road, Berkeley, CA 94720, USA}
\email{yifeiluo@lbl.gov}

\author{Aaron Meisner}
\affiliation{NSF NOIRLab, 950 N. Cherry Ave., Tucson, AZ 85719, USA}
\email{aaron.meisner@noirlab.edu}

\author{Ramon Miquel}
\affiliation{Institut de F\'{i}sica d’Altes Energies (IFAE), The Barcelona Institute of Science and Technology, Edifici Cn, Campus UAB, 08193, Bellaterra (Barcelona), Spain}
\affiliation{Instituci\'{o} Catalana de Recerca i Estudis Avan\c{c}ats, Passeig de Llu\'{\i}s Companys, 23, 08010 Barcelona, Spain}
\email{rmiquel@ifae.es}

\author{Seshadri Nadathur}
\affiliation{Institute of Cosmology and Gravitation, University of Portsmouth, Dennis Sciama Building, Portsmouth, PO1 3FX, UK}
\email{seshadri.nadathur@port.ac.uk}

\author{Nathalie Palanque-Delabrouille}
\affiliation{IRFU, CEA, Universit\'{e} Paris-Saclay, F-91191 Gif-sur-Yvette, France}
\affiliation{Lawrence Berkeley National Laboratory, 1 Cyclotron Road, Berkeley, CA 94720, USA}
\email{npalanque-delabrouille@lbl.gov}

\author{Will Percival}
\affiliation{Department of Physics and Astronomy, University of Waterloo, 200 University Ave W, Waterloo, ON N2L 3G1, Canada}
\affiliation{Perimeter Institute for Theoretical Physics, 31 Caroline St. North, Waterloo, ON N2L 2Y5, Canada}
\affiliation{Waterloo Centre for Astrophysics, University of Waterloo, 200 University Ave W, Waterloo, ON N2L 3G1, Canada}
\email{will.percival@uwaterloo.ca}

\author{Ignasi Pérez-Ràfols}
\affiliation{Departament de F\'isica, EEBE, Universitat Polit\`ecnica de Catalunya, c/Eduard Maristany 10, 08930 Barcelona, Spain}
\email{ignasi.perez.rafols@upc.edu}

\author{Graziano Rossi}
\affiliation{Department of Physics and Astronomy, Sejong University, 209 Neungdong-ro, Gwangjin-gu, Seoul 05006, Republic of Korea}
\email{graziano@sejong.ac.kr}

\author{Eusebio Sanchez}
\affiliation{CIEMAT, Avenida Complutense 40, E-28040 Madrid, Spain}
\email{eusebio.sanchez@ciemat.es}

\author{David Schlegel}
\affiliation{Lawrence Berkeley National Laboratory, 1 Cyclotron Road, Berkeley, CA 94720, USA}
\email{djschlegel@lbl.gov}

\author{Hee-Jong Seo}
\affiliation{Department of Physics \& Astronomy, Ohio University, 139 University Terrace, Athens, OH 45701, USA}
\email{seoh@ohio.edu}

\author{Joseph Harry Silber}
\affiliation{Lawrence Berkeley National Laboratory, 1 Cyclotron Road, Berkeley, CA 94720, USA}
\email{jhsilber@lbl.gov}

\author{David Sprayberry}
\affiliation{NSF NOIRLab, 950 N. Cherry Ave., Tucson, AZ 85719, USA}
\email{david.sprayberry@noirlab.edu}

\author{Gregory Tarlé}
\affiliation{University of Michigan, 500 S. State Street, Ann Arbor, MI 48109, USA}
\email{gtarle@umich.edu}

\author{Benjamin Alan Weaver}
\affiliation{NSF NOIRLab, 950 N. Cherry Ave., Tucson, AZ 85719, USA}
\email{benjamin.weaver@noirlab.edu}


\begin{abstract}
Extremely Metal-Poor Galaxies (XMPGs), defined as having metallicities below 10\% of the solar value, are considered possible local analogs to primordial systems and offer a unique window into early galaxy evolution. This study presents a large-scale search for XMPGs using data from the Dark Energy Spectroscopic Instrument DR1, systematically evaluating their resemblance to high-redshift galaxies. From a parent sample of over 14 million galaxies, we identify 656 (551 new) confirmed XMPGs and 767 (670 new) high-quality candidates via the direct $T_{\mathrm{e}}$ method. Results reveal that XMPGs follow a distinct star-forming main sequence (SFMS) that is elevated and shallower than that of the comparing star-forming galaxies. Notably, at higher stellar masses ($M_{\star} > 10^{7.5} M_{\odot}$), the XMPG SFMS converges with the sequence observed in high-redshift galaxies by James Webb Space Telescope (JWST), indicating that mature XMPGs sustain star formation rates comparable to their primordial counterparts. Furthermore, XMPGs consistently deviate below the local fundamental metallicity relation, mirroring high-redshift galaxy behavior. These findings demonstrate that XMPGs not only exhibit low metallicities but also preserve scaling relations characteristic of the early Universe, confirming their potential value as local laboratories for studying early galaxy formation processes.
\end{abstract}

\keywords{\uat{Galaxies}{573} --- \uat{Metallicity}{1031} --- \uat{High-redshift galaxies}{734} --- \uat{Emission line galaxies}{459}}


\section{Introduction} \label{sec:introduction}
Galaxy evolution is fundamentally linked to the continuous process of chemical enrichment. Metals synthesized through stellar nucleosynthesis and dispersed by supernova explosions accumulate within the interstellar medium (ISM). The resulting metallicity of a galaxy is governed by a complex equilibrium between this internal enrichment, the ejection of material through galactic outflows, and the accretion of pristine gas from the intergalactic medium \citep{2020ARA&A..58..363P}. This balance produces strong correlations between metallicity and fundamental galaxy properties such as stellar mass and star formation rate (SFR), as established by the mass-metallicity and fundamental metallicity relations \citep{2013ApJ...765..140A, 2020MNRAS.491..944C}. Consequently, these scaling relations serve as powerful diagnostics for tracing the chemical evolution history of a galaxy \citep{2018ApJ...868...89G, 2019A&ARv..27....3M}

Extremely metal-poor galaxies (XMPGs), defined as systems with gas-phase metallicities below one-tenth solar abundance ($Z < 0.1 Z_{\odot}$), represent a unique population of chemically primitive systems \citep{2000A&ARv..10....1K}. Adopting the standard solar oxygen abundance benchmark $12 + \mathrm{log(O/H)_{\odot}} = 8.69$ \citep{2009ARA&A..47..481A,2021A&A...653A.141A}, this threshold corresponds to $12 + \mathrm{log(O/H)} < 7.69$. The typical characteristics of XMPGs feature low stellar masses (median $M_{\star} \sim 5\times10^7  M_{\odot}$), elevated specific star formation rates (median $\text{sSFR} \sim 10^{-8}$ yr$^{-1}$), and , and evolutionary phases resembling those of primordial galaxies \citep{2024ApJ...961..173Z, 2025MNRAS.540..128C, 2025ApJ...988...46F}. Consequently, local XMPGs serve as valuable analogues for studying early galaxy formation processes \citep{2025MNRAS.540..128C}. Additionally, their simple chemical evolution and minimal stellar helium production make XMPGs ideal for constraining the primordial helium abundance, providing an empirical anchor that complements Big Bang Nucleosynthesis predictions and cosmic microwave background constraints \citep{2022MNRAS.510..373A}.

Despite their importance, XMPGs remain exceptionally rare. Large-scale spectroscopic surveys indicate they constitute only a few percent of galaxies in the nearby Universe \citep{2016ApJ...819..110S}, with high-redshift counterparts being even more challenging to identify due to observational limitations. Systematic searches for XMPGs have therefore focused on three primary objectives: expanding sample sizes, identifying the most metal-poor systems, and pushing detections to higher redshifts.

The pursuit of extremely metal-poor systems has revealed several notable galaxies with metallicities approaching 12 + log(O/H) $\sim$ 7.0, including KHC912-29, AGC 198691, and J0811+4730 \citep{2009ApJ...698.2014H,2016ApJ...822..108H,2018MNRAS.473.1956I,2020ApJ...898..142K,2022MNRAS.516L..81T,2024A&A...690A..28Z}. Precise metallicity determination in these low-abundance regimes remains challenging. In some cases, more robust follow-up measurements have revised initial estimates, highlighting difficulties in confirming extreme metal-poor status \citep[e.g.,][]{2020ApJ...898..142K, 2022ApJ...935..150L, 2024arXiv241204541N}.  To date, no galaxy in the local universe has been unequivocally confirmed to have a metallicity below the canonical threshold of $Z \sim 0.02 Z_\odot$. However, recent progress has shifted to higher redshifts. At $z>5$, studies utilizing the direct method (see Section \ref{sec:metallicity} for details) have acquired promising candidates with 12 + log(O/H) $\lesssim$ 7.0 \citep{2025MNRAS.540.2176C,2025arXiv250503873H,2025arXiv250717820C}. The definitive confirmation of these candidates, and thus the robust identification of XMPGs at high redshift, will require deeper spectroscopic observations and robust diagnostic checks.

While early compilations contained only a handful of confirmed XMPGs \citep{2000A&ARv..10....1K}, subsequent surveys have dramatically increased these numbers. SDSS-based studies, for example, identified dozens of XMPGs and candidates \citep[e.g.,][]{2011ApJ...743...77M,2016ApJ...819..110S}. More recently, dedicated efforts using Subaru/Hyper Suprime-Cam photometry, early Dark Energy Spectroscopic Instrument (DESI) data, and SDSS DR17 data have yielded hundreds of confirmed XMPGs and candidates \citep{2020ApJ...898..142K,2024ApJ...961..173Z,2025MNRAS.540..128C}. Establishing a robust connection between these low-redshift XMPGs and galaxies in the early universe requires statistically significant and unified samples to overcome observational biases and reveal fundamental evolutionary links.

In this paper, we utilize the extensive spectroscopic data from the DESI Data Release 1 (DR1) to assemble the largest catalog of XMPGs to date and examine their possible links to high-redshift galaxies. The structure of this paper is as follows. Section \ref{sec:data} outlines the DESI dataset and sample selection. Section \ref{sec:metallicity} details the method for metallicity estimation. In Section \ref{sec:xmpgs}, we introduce the identified XMPG sample, analyze their scaling relations including the star formation main sequence and fundamental metallicity relation, and discuss potential connections to high-redshift systems. Finally, Section \ref{sec:summary} provides a concluding summary of our findings.

\section{DESI Data and Sample Selection} \label{sec:data}
\subsection{DESI overview}
As the first Stage-IV dark-energy experiment, DESI is designed to map the large-scale structure of the universe out to redshifts of $z \sim 3.5$ \citep{2006astro.ph..9591A, 2013arXiv1308.0847L, 2016arXiv161100036D, 2016arXiv161100037D}. Installed on the Mayall 4-meter telescope at Kitt Peak National Observatory in Arizona, USA \citep{2022AJ....164..207D}, it consists of 10 petals that collectively provide a $3.2^{\circ}$-diameter field of view \citep{DESI2016b.Instr,Corrector.Miller.2023}. Each petal is equipped with optical fibers that project to $1.5 \arcsec$ on the sky, allowing for rapid reconfiguration and enabling the simultaneous acquisition of up to 5000 spectra \citep{FiberSystem.Poppett.2024}. The fibers feed ten spectrographs, each with three wavelength channels, covering $3600 - 9800 \mathrm{\AA}$ at a resolution of $\lambda / \Delta \lambda \sim 2000 - 5000$. This resolution is sufficient to resolve the $\mathrm{[O\ II]}\lambda \lambda 3726,3729$ doublet at redshifts $z > 0.49$. 

The targets for the DESI spectroscopic survey are selected based on photometric data from the DESI Legacy Imaging Surveys \citep[hereafter Legacy Surveys,][]{2019AJ....157..168D}. The Legacy Surveys combine data from three public projects:the Dark Energy Camera Legacy Survey (DECaLS), the Beijing–Arizona Sky Survey (BASS) \citep{2017PASP..129f4101Z}, and the Mayall z-band Legacy Survey (MzLS). Their respective 5$\sigma$ point-source limiting magnitudes in the $g$, $r$, and $z$ bands are 23.95, 23.54, and 22.50 mag for DECaLS, and 23.65, 23.08, and 22.60 mag for the combined BASS/MzLS data \citep{2019AJ....157..168D}. Additional infrared photometry in four bands is incorporated from the NEOWISE mission \citep{2014ApJ...792...30M}.

DESI began a five-month survey validation (SV) phase in December 2020 to assess instrument performance, target selection, data processing pipelines, and scientific data quality \citep{2024AJ....167...62D}. The main spectroscopic survey commenced in May 2021 and is scheduled to operate for five years, covering 14,000 deg$^2$ \citep{SurveyOps.Schlafly.2023}, and ultimately collecting spectra for approximately 50 million galaxies and quasars, plus 25 million Milky Way stars \citep{2025arXiv250314745D}.  DESI targets include luminous red galaxies (LRGs, $0.4 < z < 1.1$), emission-line galaxies (ELGs, $0.6 < z < 1.6$), quasars (QSOs, $0.9 < z < 4$), bright galaxy survey (BGS) galaxies ($0 < z < 0.6$), and Milky Way survey (MWS) stars \citep{2023AJ....165...50M}. Beyond its primary targets, DESI employs spare fibers to conduct secondary surveys, such as the LOW-Z Survey \citep{2023ApJ...954..149D} that is designed to observe faint dwarf galaxies at low redshift.

\begin{figure*}[htbp!]
\plotone{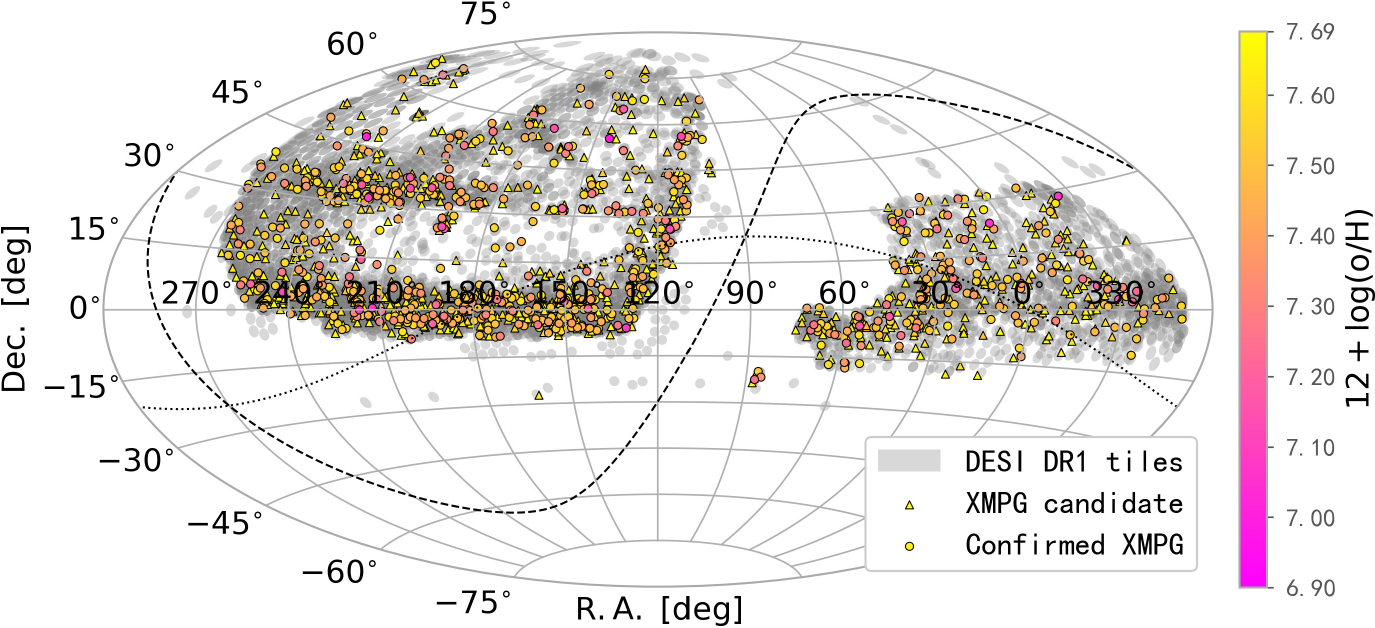}
\caption{Sky coverage and XMPG sample in DESI DR1. The gray background indicates the observed sky areas in DR1, with darker shades corresponding to higher coverage density. Confirmed XMPGs are shown as circles, and XMPG candidates as triangles (see Section \ref{sec:xmpgs}), which are color-coded by metallicity. The dashed and dotted lines mark the Galactic and ecliptic planes, respectively. \label{fig:Iron_XMPG_coord}}
\end{figure*}

Data from the SV phase were released as the DESI Early Data Release (EDR) in June 2023 \citep{2024AJ....168...58D}, containing about 1.7 million well-calibrated spectra, including over 1.1 million galaxy spectra. In March 2025, DESI DR1 was made public \citep{2025arXiv250314745D}, incorporating the first 13 months of main-survey observations along with uniformly reprocessed SV and EDR data. DR1 includes approximately 13.1 million galaxies, 1.6 million quasars, and 4 million stars. Early results from DESI have already placed strong constraints on dark energy \citep{DESI2024.VII.KP7B, 2025PhRvD.112h3515A}. The sky coverage of the survey is illustrated in Figure \ref{fig:Iron_XMPG_coord}. This study makes use of the galaxy sample from DESI DR1, benefiting from its homogeneous spectroscopic dataset and extensive multi-wavelength ancillary information.

The DESI data processing employs an extensive spectroscopic reduction pipeline \citep{Spectro.Pipeline.Guy.2023}, a template-fitting pipeline to derive classifications and redshifts for each targeted source \citep[Redrock;][]{Redrock.Bailey.2025}. Building on these foundational data products, the DESI collaboration produces Value-Added Catalogs (VACs) for each data release. These VACs supplement the basic spectral and redshift measurements with derived quantities optimized for scientific analysis.  This study utilizes the Stellar Mass and Emission Line Catalog (Mass\_EMLines \footnote{\url{https://data.desi.lbl.gov/doc/releases/dr1/vac/stellar-mass-emline/}}) from the DESI DR1 VACs. The catalog provides derived physical properties for galaxies with reliable redshifts (ZWARN $==$ 0 \& SPECTYPE $==$ ``GALAXY"), including emission-line measurements and stellar masses. Stellar masses $M_{\mathrm{\star}}$ and their associated uncertainties are derived from the Bayesian posterior probability distributions of the spectral energy distribution (SED) fitting with CIGALE \citep{2019A&A...622A.103B, 2020MNRAS.491..740Y, 2022ApJ...927..192Y}. The fitting combines broad-band $g$, $r$, $z$, $W1$, and $W2$ photometry from the Legacy Surveys with DESI spectrophotometry. Simple aperture corrections for the fiber spectra are applied by comparing the total $r$-band photometric flux to the spectrophotometric flux within the same band for each target \citep[see ][for more details]{2024ApJ...961..173Z}. The final catalog contains 14,706,085 entries, corresponding to 14,183,563 are unique galaxies.

\subsection{Star-forming galaxy sample} \label{subsec:data selection}

Gas-phase metallicities are determined using the electron temperature ($T_\mathrm{e}$) method \citep{1984ASSL..112.....A}, which requires detection of the weak auroral line $\mathrm{[O\ III]} \lambda 4363$. Since this line is undetectable in a substantial fraction of galaxies, a subsample is selected from the Mass\_EMLines catalog where $T_\mathrm{e}$ -based metallicities can be computed. The selection imposes the following criteria on emission lines $\mathrm{[O\ II]} \lambda \lambda 3726,3729$, $\mathrm{[O\ III]} \lambda 4363$, $\mathrm{[O\ III]} \lambda \lambda 4959,5007$, and $\mathrm{H}\beta$ to minimize false detection: 
full width at half maximum (FWHM) in the rest frame $> 1$ {\AA}, signal-to-noise ratio (S/N) $>$ 3, and central wavelength deviation from theoretical values $< 0.5$ {\AA}. Relatively stringent constraints on the central wavelengths of emission lines are applied to mitigate spurious detections. This is primarily intended to prevent false line identifications caused by noise-induced wavelength shifts and erroneous fits with abnormally large line widths in low signal-to-noise (S/N) spectra, particularly for the faint auroral line [O III]$\lambda$4363. However, it should be noted that this process may also exclude some genuine galaxies whose emission lines are shifted due to galactic outflows. For extinction correction, either $\mathrm{H}\alpha$ or $\mathrm{H}\gamma$ must satisfy these same criteria. Broad-line active galactic nuclei (AGNs, typically with FWHM $> 1,000$ km/s; \citealt{2001AJ....122..549V}) are excluded by imposing a stricter constraint of  $\mathrm{FWHM(H}\beta) < 500$ km/s (equivalent to about 8.1 $\mathrm{\AA})$. Furthermore, stellar masses are required to satisfy two quality criteria: $\mathrm{log}(M_{\star}/M_{\odot}) > 5$ (to exclude potential substructures of large nearby galaxies) and an associated uncertainty of $\sigma_{\mathrm{log}M_{\star}} < 0.4$. The latter serves as a conservative threshold to exclude galaxies with clearly unreliable SED fits. These criteria yield a sample of 9,579 galaxies suitable for $T_\mathrm{e}$-based metallicity calculations, with redshift limited to $z \leq 0.96$ to ensure $\mathrm{[O\ III]} \lambda 5007$ line remains within the spectral coverage.

Intrinsic gas-phase extinction is derived to correct observed emission-line fluxes using the color excess $\mathrm{E(B-V)}$ computed from Balmer decrements ($\mathrm{H}\alpha / \mathrm{H}\beta$ or $\mathrm{H}\gamma / \mathrm{H}\beta$). Assuming Case B recombination with electron density $N\mathrm{_{e}=100 cm^{-3}}$ and $T\mathrm{_{e}=10,000 K}$, the intrinsic ratios are $\mathrm{H}\alpha / \mathrm{H}\beta = 2.86$ and $\mathrm{H}\beta / \mathrm{H}\gamma = 2.137$ \citep{2012MNRAS.419.1402G, 2024ApJ...961..173Z}. Balmer lines used for extinction estimation must have S/N $> 3$, with preference for $\mathrm{H}\alpha / \mathrm{H}\beta$ when available ($z \lesssim 0.49$). In some cases, the observed Balmer line ratios are found to be lower than the theoretical intrinsic values, leading to a formally negative calculated $E(B-V)$. This can result from genuine physical or from observational uncertainties and systematic biases in flux measurement or spectral calibration. For the 2,859 galaxies in our sample where this occurs, we set $E(B-V) = 0$, a standard practice that interprets non-positive values as indicative of negligible dust extinction for subsequent analysis. Extinction corrections adopt the \citet{1989ApJ...345..245C} law. Uncertainties are estimated via 1000 Monte Carlo realizations of flux perturbations, with final $\mathrm{E(B-V)}$ taken as the median and uncertainty as half the 16th–84th percentile range.

With the emission-line fluxes extinction-corrected, narrow-line AGNs are excluded using the diagnostic diagrams from  \citet{2024ApJ...961..173Z}: $\mathrm{[O\ III]} \lambda 5007 / \mathrm{H} \beta$ versus $\mathrm{[N\ II]} \lambda 6583 / \mathrm{H} \alpha$, $\mathrm{[O\ III]} \lambda 5007 / \mathrm{H} \beta$ versus $\mathrm{[S\ II]} \lambda \lambda 6716,6731 / \mathrm{H} \alpha$, and $\mathrm{[O\ III]} \lambda 5007 / \mathrm{H} \beta$ versus $\mathrm{[O\ II]} \lambda \lambda 3726,3729 / \mathrm{H} \beta$. This yields a final sample of 9,372 star-forming galaxies.

\section{Metallicity Estimation Method} \label{sec:metallicity}

Among various metallicity determination methods \citep{2008ApJ...681.1183K, 2019ARA&A..57..511K}, the $T_{\mathrm{e}}$ (direct) method \citep{1984ASSL..112.....A} is widely regarded as the most reliable technique \citep{2018ApJ...869...15G, 2024ApJ...961..173Z}. This approach derives the electron temperatur  by utilizing the flux ratio of a temperature-sensitive auroral line (e.g., $\mathrm{[O\ III]} \lambda 4363$) to the stronger collisionally excited lines ($\mathrm{[O\ III]} \lambda \lambda 4959,5007$). A higher ratio of $\mathrm{[O\ III]} \lambda 4363$/$\mathrm{[O\ III]} \lambda \lambda 4959,5007$ indicates a higher electron temperature, which is typically found in low-metallicity environments where cooling is less efficient. Subsequently, this electron temperature is used to derive the ionic abundances of oxygen.

The electron temperature $T_{\mathrm{e}} \mathrm{(O\ III)}$ is determined using the \texttt{getCrossTemDen} or \texttt{getTemDen} methods in PyNeb\footnote{\url{https://research.iac.es/proyecto/PyNeb//}}, together with the electron density $N_e$. When feasible, $T_{\mathrm{e}} \mathrm{(O\ III)}$ and $N_\mathrm{e}$ are derived iteratively via the \texttt{getCrossTemDen} method. For this iterative process, $N_e$s primarily calculated from the $\mathrm{[S\ II]}$ or $\mathrm{[O\ II]}$ doublet ratios. In cases where neither doublet is detectable, a default value of $N_\mathrm{e} = 100$ cm$^{-3}$ is adopted, following established practices in previous studies \citep{2018ApJ...869...15G, 2020ApJ...898..142K, 2024ApJ...961..173Z}. This assumption is valid because $T_{\mathrm{e}} \mathrm{(O\ III)}$ shows negligible dependence on $N_e$ at densities below $10^4 \mathrm{cm^{-3}}$ \citep{2014ApJ...780..122L}. Under such default density conditions, $T_{\mathrm{e}} \mathrm{(O\ III)}$ is computed using the \texttt{getTemDen} method. It is noted that the standard methods (\texttt{getCrossTemDen} and \texttt{getTemDen}) are subject to an upper temperature limit of $\sim 25,000$K \citep{2020ApJ...898..142K}. These limitations can prevent the derivation of electron temperatures for certain galaxies. For the 1,781 galaxies in the sample where these conventional methods failed,  $T_{\mathrm{e}} \mathrm{(O\ III)}$ is computed using the  PyNeb \texttt{getEmissivity} method. This approach supports temperatures up to 200,000 K \citep{1999ApJS..123..311A, 2015A&A...573A..42L},  with an assumed $N\mathrm{_{e} = 100 cm^{-3}}$ for these calculations. The electron temperature for the low-ionization zone, O\ II, is subsequently estimated from the empirical relation: $T_{\mathrm{e}} \mathrm{(O\ II)} = 0.7 \times T_{\mathrm{e}} \mathrm{(O\ III)} + 3000$, as used by \citet{1986MNRAS.223..811C} and \citet{2020ApJ...898..142K}.

The ionic abundances are calculated using the \texttt{getIonAbundance} method in PyNeb. Specifically, the relative abundance $\mathrm{O}^{+}/\mathrm{H}^{+}$ is derived from the line ratio $\mathrm{[O\ II]}\lambda \lambda3726,3729/\mathrm{H}\beta$ for $\mathrm{O}^{+}$, while $\mathrm{O}^{++}/\mathrm{H}^{+}$ is derived from $\mathrm{[O\ III]}\lambda \lambda4959, 5007/\mathrm{H}\beta$. The total oxygen abundance is then calculated under the assumption that the contribution from higher ionization states (e.g.,  $\mathrm{O}^{+++}$) is negligible \citep{2021ApJ...922..170B}, following the expression:
\begin{equation}
\frac{\mathrm{O}}{\mathrm{H}}=\frac{\mathrm{O}^{+}}{\mathrm{H}^{+}} + \frac{\mathrm{O}^{++}}{\mathrm{H}^{+}},
\label{eq:O_ion}
\end{equation}
This resulting value is subsequently converted to the standard metallicity scale, denoted as 12 + log(O/H). To estimate the uncertainties in the derived quantities, a Monte Carlo approach is employed, involving 1000 realizations of random perturbations to the emission line fluxes according to their measurement errors. The final uncertainty for each parameter is taken as half of the 16th to 84th percentile range of the resulting distribution.
 
\begin{figure*}[ht!]
\plotone{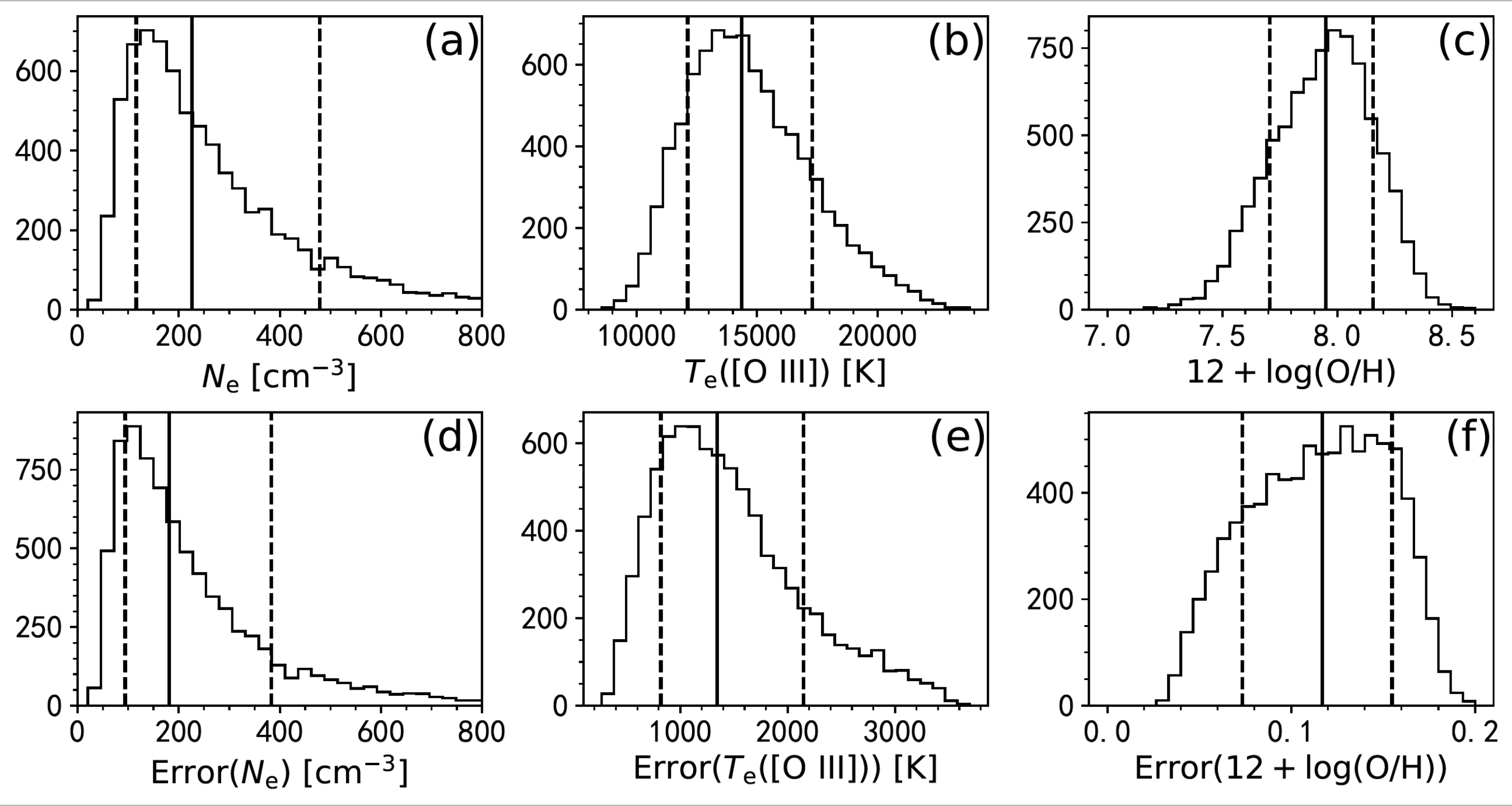}
\caption{Distributions for our $\mathrm{[O\ III]} \lambda 4363$-detected star-forming galaxies: (a)$N_\mathrm{e}$, (b) $T_{\mathrm{e}} \mathrm{([O\ III])}$, and (c) metallicity, and the corresponding errors in (d-f). Vertical lines indicate medians.
\label{fig:oiii_galaxy_properties}}
\end{figure*}

Figure \ref{fig:oiii_galaxy_properties} shows distributions of derived parameters, including $N_\mathrm{e}$, $T_{\mathrm{e}}$, metallicity, and their associated errors for our $\mathrm{[O\ III]} \lambda 4363$-detected galaxy sample. Electron densities are measured for 3,999 galaxies using $\mathrm{[S\ II]} \lambda 6731 / \mathrm{[S\ II]} \lambda 6716$ and 3,167 using $\mathrm{[O\ II]} \lambda 3726 / \mathrm{[O\ II]} \lambda 3729$, with median $N_\mathrm{e} = 175$ cm$^3$ (median error of 186 cm$^3$). $T_{\mathrm{e}} \mathrm{([O\ III])}$ is obtained for 9,314 galaxies (median: 14,454 K; median error: 1,375 K). Metallicities are derived for 9,292 galaxies (median: 7.94 dex; error: 0.12 dex), with 1,475 sources below 12 + log(O/H) = 7.69.  

To verify the robustness of the temperature determinations, the consistency between $T_\mathrm{e}([\mathrm{O\ III}])$ values derived from different methods and density assumptions is examined. For the 2,126 galaxies whose temperatures were derived assuming a fixed density of $N_\mathrm{e} = 100$ cm$^{-3}$, the median deviations are 3.45K (standard deviation: 3.93 K) when adopting $N_\mathrm{e}=10$ cm$^{-3}$ and $-29.01$ K (standard deviation: 13.87K) when adopting $N_\mathrm{e}=10$ cm$^{-3}$. Additionally, for the 7,591 galaxies with temperatures derived from the standard methods of \texttt{getCrossTemDen} and \texttt{getTemDen}, comparison with the \texttt{getEmissivity} method yields a median deviation of 2.70K (standard deviation: 12.96K). The variations introduced by different methods and assumptions are negligible compared to the typical measurement uncertainty in $T_\mathrm{e}([\mathrm{O\ III}])$, confirming the robustness of our temperature determinations.

\section{Our XMPG Sample and Connections to High-$z$ Galaxies} \label{sec:xmpgs}

\subsection{The XMPG sample} \label{subsec:properties}

From the initial 1,475 sources with $12 + \mathrm{log(O/H)} < 7.69$ identified in Section \ref{sec:metallicity}, visual inspection excluded 52 objects contaminated by foreground stars, instrumental artifacts, or determined to be substructures of large nearby galaxies. The remaining 1,423 sources were classified based on their metallicity uncertainties: 656 galaxies where the upper bound of the metallicity measurement (including 1$\sigma$ uncertainty) falls below 7.69 are designated as confirmed XMPGs, while the remaining 767 sources are classified as XMPG candidates. A review of pre-DESI literature identifies 351 previously reported XMPGs as summarized by \citet{2024ApJ...961..173Z}, alongside 223 XMPGs from early DESI data in the same study. This work utilizes the newer DESI DR1 dataset reduced with an updated pipeline and incorporates revised methods for deriving $Te$ and metallicity. Consequently, some XMPGs identified in the EDR may not meet the selection criteria in DR1 due to  methodological minor adjustments and measurement uncertainties. Our sample shares 27 objects with the pre-DESI literature and 180 objects with the EDR sample. In total, we newly identify 1,221 XMPGs (551 confirmed and 670 candidates), substantially expanding the known population of these rare systems.

\begin{figure*}[ht!]
\plotone{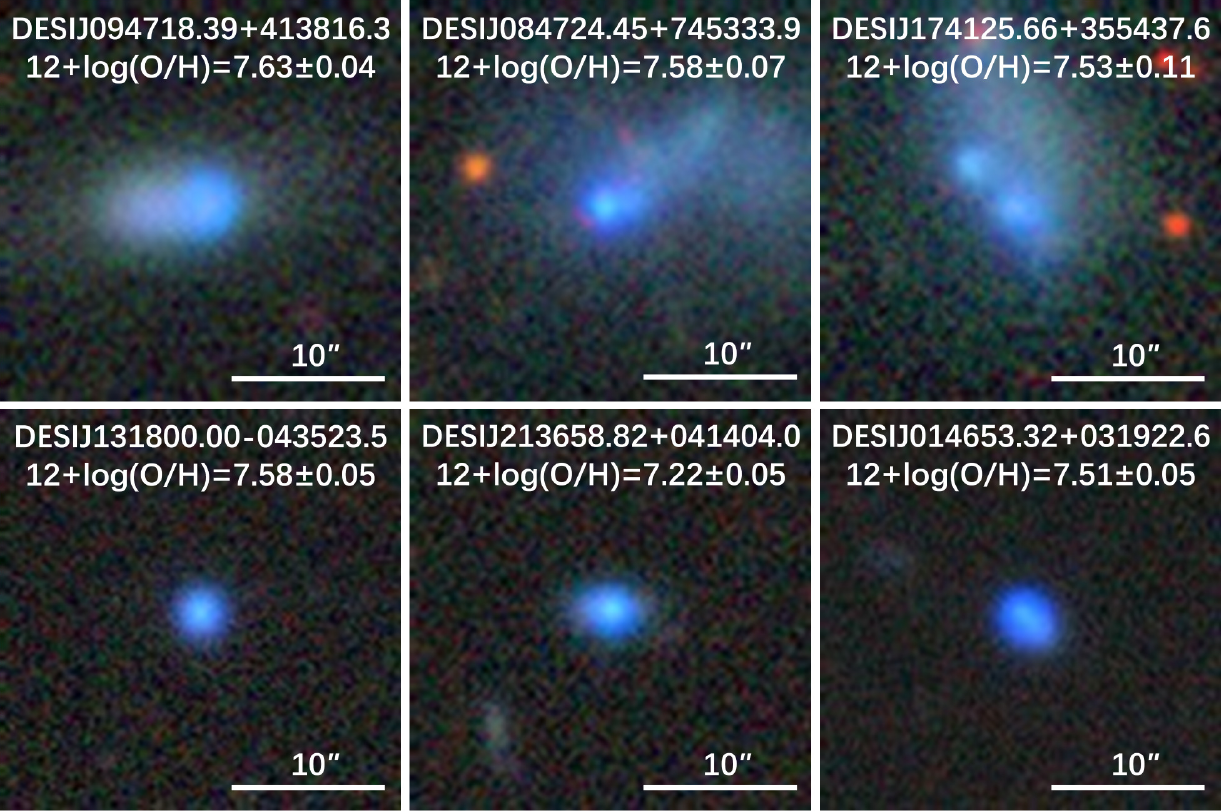}
\caption{Representative examples of low-redshift XMPGs from our sample. The upper row shows three diffuse systems, while the bottom row displays three compact systems. Each panel includes the target name and metallicity (12 + log(O/H)).
\label{fig:typical_XMPGs}}
\end{figure*}

Figure \ref{fig:typical_XMPGs} presents representative examples of nearby XMPGs, highlighting their morphological diversity. While some systems appear as blue, compact objects, others exhibit complex structures with blue, metal-poor clumps superimposed on redder or more diffuse host galaxies. This diversity suggests varying origins for the metal-poor gas, that would be a scenario to be explored further with multi-wavelength data and follow-up spectroscopy. For systems with complex morphologies, the photometric data may correspond primarily to the brightest, central substructure, while diffuse emission and fainter clumps can be missed. Consequently, this may lead to a systematic underestimation of the total galactic flux, and thus of the integrated luminosity and stellar mass for the whole system. Nonetheless, the properties of the metal-poor components themselves remain reasonably constrained.

\begin{figure*}[ht!]
\plotone{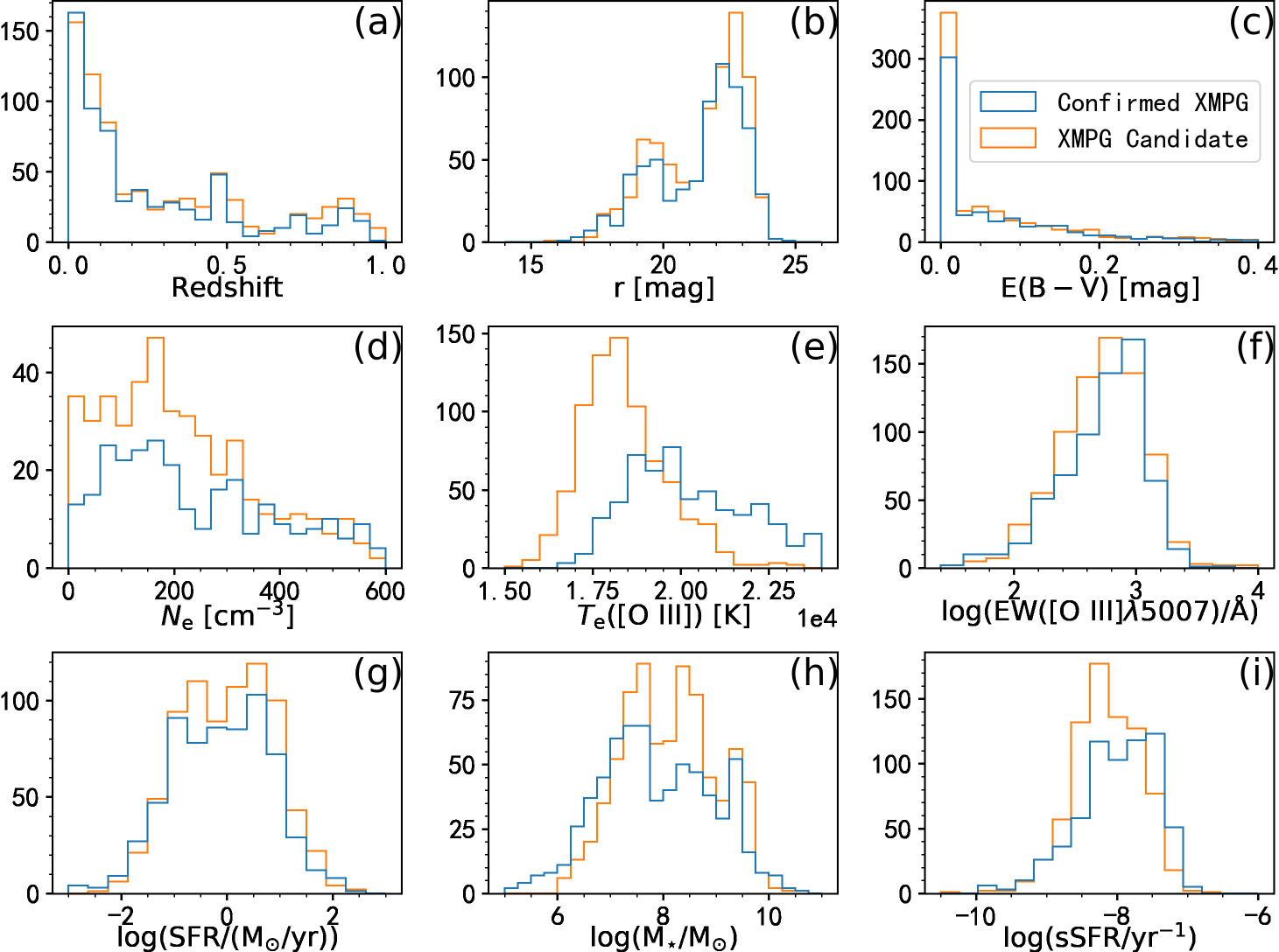}
\caption{Parameter distributions for confirmed XMPGs (blue) and XMPG candidates (orange) from DESI DR1: (a) redshift, (b) $r$-band magnitude, (c) $\mathrm{E(B-V)}$, (d) electron density ($N_{\mathrm{e}}$) without assumed values, (e) electron temperature of [O III] ($T_\mathrm{e}$([O III])), (f) log(EW( $\mathrm{[O\ III]} \lambda5007$)), (g) log(SFR), (h) log($M_*$), and (i) log(sSFR).
\label{fig:XMPG_properties}}
\end{figure*}

The physical properties of the XMPG sample are presented in Figure \ref{fig:XMPG_properties}. The SFRs are measured from hydrogen recombination lines \citep{1998ARA&A..36..189K} using:
\begin{equation}
\mathrm{SFR}=7.9 \times 10^{-42} L(\mathrm{H} \alpha) ~M_{\odot} \mathrm{yr}^{-1} \mathrm{erg}^{-1} \mathrm{~s},
\label{eq:SFR}
\end{equation}
where $L(\mathrm{H} \alpha) = 4\pi d^2_L F(\mathrm{H} \alpha)$ is the $\mathrm{H} \alpha$ luminosity, $F(\mathrm{H} \alpha)$ is the extinction-corrected $\mathrm{H} \alpha$ flux, and $d_L$ is the luminosity distance. This calculation assumes a Salpeter initial mass function \citep[IMF,][]{1955ApJ...121..161S}.  The SFR is derived primarily from the $\mathrm{H} \alpha$ emission line. For sources at $z > 0.49$ or with a S/N $< 3$ in $\mathrm{H} \alpha$, we use the $\mathrm{H} \beta$ instead, assuming an intrinsic flux ratio of $F(\mathrm{H} \alpha)/F(\mathrm{H} \beta) = 2.86$. Equation (\ref{eq:SFR}) has a systematic uncertainty of approximately $30\%$ (dominated by calibration uncertainties), which is incorporated into the total SFR error budget. Aperture corrections are applied using the same scaling factors as those adopted for stellar mass derivation in the Mass\_EMLines catalog. Finally, the SFRs are converted to a Chabrier initial mass function (IMF) by multiplying by a factor of 0.63 \citep{2024MNRAS.532.2832L}, ensuring consistency with the stellar mass estimates \citep{2003PASP..115..763C}. Additionally, we tested the robustness of our results by applying the metallicity-dependent SFR calibration from \citet{2018ApJ...869...92R}. This alternative calibration was found to have a negligible impact on the main results about the star-forming main sequence and fundamental metallicity relation, as presented in Sections \ref{subsec:SFMS} and \ref{subsec:FMR}. It is worth noting that because the aperture correction method assumes proportionality in flux across different wavelength bands for all galactic regions, if the fiber target corresponds to a region with strong Balmer lines within the galaxy, this will lead to deviations in the Balmer-line luminosity of the galaxy and further result in the overestimation of the SFR.

As shown in Figure \ref{fig:XMPG_properties}, the XMPG sample is limited to $z<1$ due to the requirement of detecting the $\mathrm{[O\ III]} \lambda5007$ line. The redshift distribution shows a slight peak near $z\sim0.5$ arises, which is primarily contributed by the ELG targets.  The $r$-band magnitude distribution in Figure \ref{fig:XMPG_properties}(b) exhibits a bimodal structure: one peak around $r\sim$19--20 is dominated by BGS targets, while XMPGs from other target classes concentrate at fainter magnitudes around $r\sim$22--23. The extinction distribution in this figure gives the median $E(B-V) \approx 0.03$, indicating low dust content. Confirmed XMPGs, which account for the metallicity measurement uncertainties, exhibit lower metallicities than the candidates, corresponding to higher $T_e$. This elevated electron temperature contributes to the large equivalent width of the [O III] emission line. The stellar mass and SFR distributions reveal that most XMPGs are dwarf galaxies ($M_{\star} < 10^9 M_{\odot}$) with intense star formation, having sSFR typical of starburst galaxies. The median stellar mass is about $9.65 \times 10^7$ M$_\odot$ and median sSFR is about $8.62 \times 10^{-9} \mathrm{yr}^{-1}$. Properties of confirmed XMPGs and candidates are listed in Table \ref{tab:conformed_XMPG} and Table \ref{tab:XMPG_candidate} in the Appendix, respectively.

\subsection{The Most Metal-Poor Galaxy} \label{subsec:most MPG}

The most metal-poor galaxy identified in this work is DESIJ093402.37+551423.2 (DESI TARGETID: 39633324993414901) at redshift z=0.0026. This source corresponds to the southeast component of the well-known metal-poor blue compact dwarf galaxy I Zw 18 \citep{1972ApJ...173...25S, 1997ApJ...487L..37I}. While cataloged separately in this paper, it is not an independent system; therefore, in relevant analyses, we treat it as part of the larger I Zw 18 complex. Located at a distance of $18.2 \pm 1.5$ Mpc \citep{2007ApJ...667L.151A}, I Zw 18 as a whole is often considered a local analog of high-redshift dwarf galaxies due to its extremely low metallicity, low stellar mass \citep[\(10^6 - 10^7 M_{\star}\)][]{2020A&A...641A.168N, 2024AJ....168...23H} and high sSFR \citep[\(10^{-8} - 10^{-7} yr^{-1} \),][]{2025ApJ...992...48H}. The system is currently undergoing intense star formation, likely triggered by interaction with its companion galaxy \citep{2024A&A...689A.146B, 2025arXiv250204965B}. However, some recent studies suggest that I Zw 18 may not be a young system but rather an older galaxy experiencing a recent starburst episode \citep{2024A&A...689A.146B}.

In our work, DESIJ093402.37+551423.2 exhibits a metallicity of 12+log(O/H)=$6.99 \pm 0.07$ approaching the current lower metallicity limit for low-redshift galaxies ($\sim 2\%$ of solar). This value is slightly lower than some previously reported measurements of $\sim 7.2$ \citep{2016MNRAS.459.2992K, 2025ApJ...990..111R}, and is close to the value of $7.07 \pm 0.05$ measured by \cite{2026arXiv260108722Z}. With a stellar mass of $M_{\star}=(2.78 \pm 0.98) \times 10^5 M_{\odot}$ and an SFR of 0.0093 $M_{\odot} / \mathrm{yr}$, this object shows a high sSFR of $3.36 \times 10^{-8} \mathrm{yr}^{-1}$, consistent with a starburst dwarf system. 

\begin{figure*}
\centering
	\includegraphics[width=0.8\textwidth]{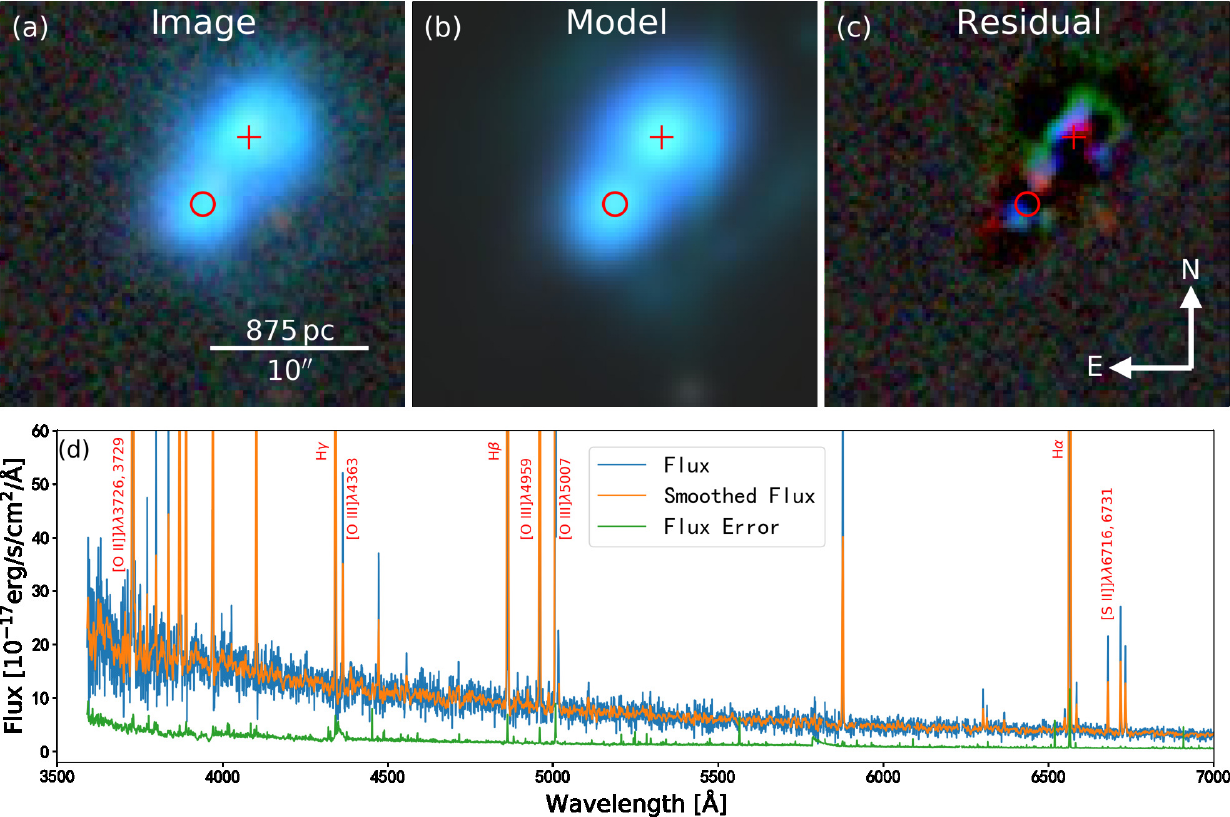}
    \caption{The most metal-poor galaxy in this work (DESI J093402.37+551423.2), a component of the well-known system I Zw 18. Panels (a)-(c) show its Legacy Surveys DR9 data and  morphological modelling: (a) color image, (b) best-fit model, and (c) residual map after model subtraction. The red circle marks the 1.5${\arcsec}$ DESI fiber aperture centered on this object; the red cross marks the northwestern component of I ZW 18 (designated as DESIJ093402.01+551427.6). Panel (d) presents the optical spectrum with key emission lines labelled. \label{fig:DESIJ093402.37+551423.2}}
\end{figure*}

Figure \ref{fig:DESIJ093402.37+551423.2} (a)-(c) present the the color image and morphological model and corresponding residual map from Legacy Surveys DR9 \citep{2025ascl.soft02024L}. Model photometry yields magnitudes of $g=$17.59, $r=$17.53, and $z=$18.22, with an absolute $r$-band magnitude of $M_r=$-13.77. The half-light radius is $R_e = $1.64 $\pm 0.0025$ arcsec, equivalent to 143.35 $\pm 11.82$ pc at the galaxy distance. The optical spectrum in Panel (d) shows strong emission lines and a continuum dominated by massive young stars, confirming intense ongoing star formation.

A neighboring object, located northwest of DESI J093402.37+551423.2 and marked with a red cross in Figure \ref{fig:DESIJ093402.37+551423.2}, corresponds to the northwest component of I Zw 18 designated as DESIJ093402.01+551427.6 in DESI, but not included in DESI DR1). This companion is also metal-poor, with reported metallicities in the range 12+log(O/H)=$\sim 7 - 7.15$ \citep{2016MNRAS.459.2992K, 2022ApJS..261...31B, 2025ApJ...990..111R}. Among them, the lowest recorded metallicity of this object is 12 + log(O/H) $= 6.98 \pm 0.01$, measured by \cite{2022ApJS..261...31B}, which is consistent with our measurement of the southeast component. The lowest recorded value of 6.98$\pm$0.01 \citep{2022ApJS..261...31B}, is consistent with our measurement of the southeast component. The two systems are separated by 0.47 kpc in projection. DESIJ093402.01+551427.6 has optical magnitudes of $g=$16.31, $r=$16.43, $z=$17.03, with $M_r$=-14.87 and a half-light radius of $R_e$=161.81 $\pm 13.34$ pc. The stellar mass is $M_\star = (1.12 \pm 0.29) \times 10^6 M_{\odot}$, approximately four times that of the southeast component. The close proximity, similar blue colors, and comparable stellar masses suggest a possible common origin for the two components. As noted by \citet{2024A&A...689A.146B}, the intense star formation in I Zw 18 may result from interaction with another companion galaxy. Furthermore, other studies have indicated that the SE and NW components of I Zw 18 may themselves be two merging galaxies \citep{2024arXiv241204541N}.

\subsection{Star-forming main sequence} \label{subsec:SFMS}

The star formation main sequence (SFMS) represents a fundamental empirical relation between $M_{\star}$ and SFR in star-forming galaxies (SFGs). Since its initial identification, this relation has served as a key benchmark for understanding galaxy evolution: at a given cosmic epoch, most SFGs cluster tightly around a well-defined trend, where SFR increases monotonically with $M_{\star}$, exhibiting an intrinsic scatter $< 0.3$ dex \citep{2007ApJ...660L..43N}. A defining characteristic of the SFMS is its strong redshift evolution: the normalization (SFR at fixed $M_{\star}$) decreases systematically with cosmic time \citep{2014ApJS..214...15S, 2023MNRAS.519.1526P, 2024A&A...683A.184G}. This evolution is linked to processes such as gas depletion, feedback from stars and AGNs, and cosmic structure formation, making the SFMS a critical tool for constraining galaxy formation models.

\begin{figure*}[ht!]
\plotone{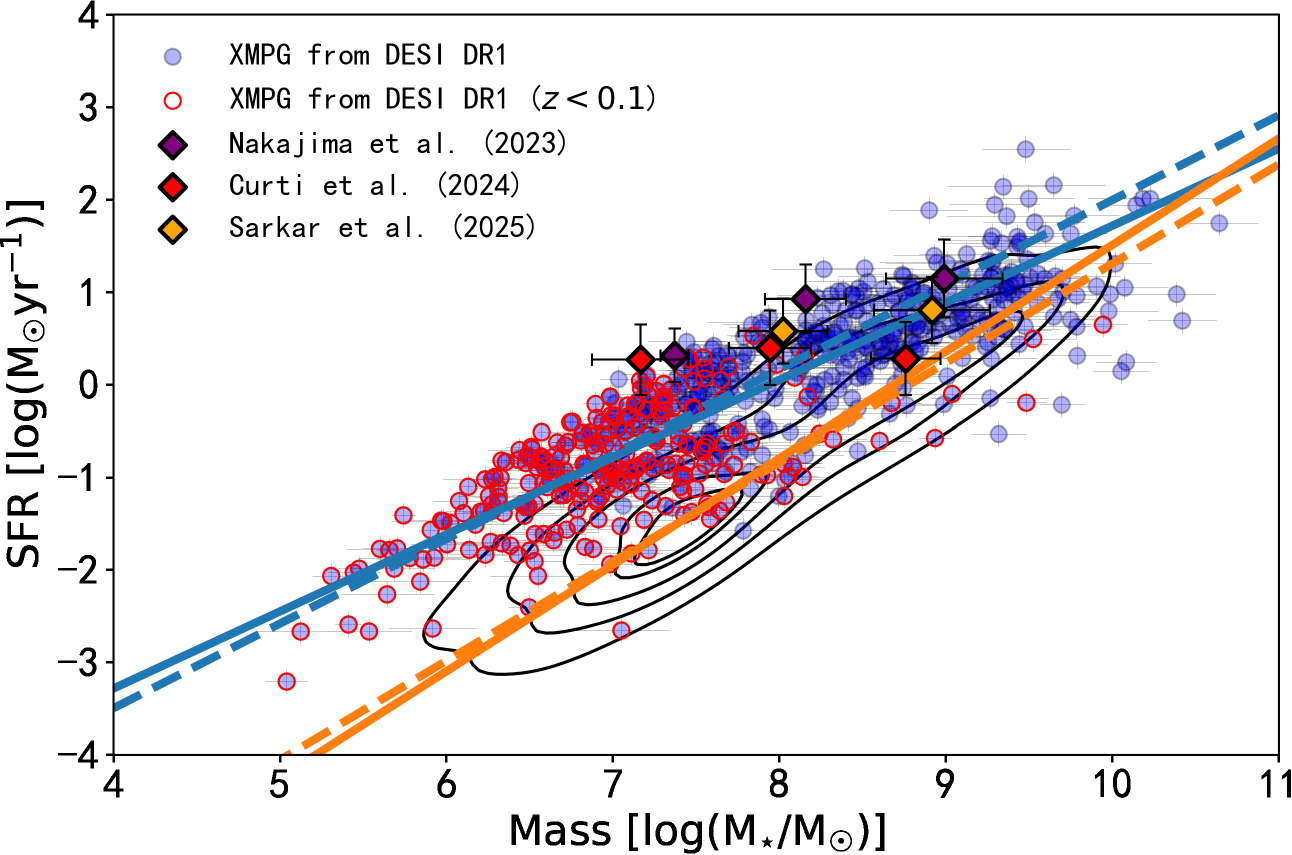}\caption{Comparison of the SFMS for our XMPG sample (blue filled circles; low-redshift ($z<0.1$) subsets highlighted with red edges), a control sample (black contours), and high-redshift JWST galaxies. JWST galaxies are binned by stellar mass ($M_{\star}<10^{7.5}M_{\odot}$,  $10^{7.5}M_{\odot}<M_{\star}<10^{8.5}M_{\odot}$, and $M_{\star}>10^{8.5}M_{\odot}$) and shown as diamonds: purple for galaxies from \cite{2023ApJS..269...33N} (\(3.8 < z < 8.7\)), red for \cite{2024A&A...684A..75C} (\(3.0 < z < 8.5\)), and orange for \cite{2025ApJ...978..136S} (\(4.0 < z < 9.4\)). For these JWST galaxies, the bins with fewer than 5 samples are excluded. The blue and orange lines represent the linearly fitted SFMS for the XMPGs and the control sample, respectively. Solid and dashed lines show the best-fit SFMS for the full and low-redshift samples, respectively. \label{fig:XMPG_VS_Control_Sample}}
\end{figure*}

To compare the SFMS of XMPGs with that of typical galaxies, a control sample was constructed. This control sample was selected to have S/N($[\mathrm{O III}]\lambda4363) <1$ ensuring that no low‑metallicity systems were inadvertently included. All other selection criteria including emission-line quality and AGN screening were identical to those used for the XMPG sample. Furthermore, the control sample was constructed to match the 656 confirmed XMPGs in $M_\star$ and redshift distributions, with an identical number of sources. This matching approach, while not eliminating all sample selection effects, enables a controlled comparison of their SFMS relations. Linear fits are performed to characterize the SFMS for both samples. Figure \ref{fig:XMPG_VS_Control_Sample} compares the resulting relations, revealing two key distinctions of the XMPG sequence: (1) a higher normalization at the low-mass end, indicating systematically elevated SFRs at fixed stellar mass compared to normal SFGs; and (2) a shallower slope, signifying a weaker dependence of SFR on stellar mass. These results demonstrate that XMPGs follow a fundamentally distinct SFMS characterized by enhanced star formation activity and efficiency that is less dependent on stellar mass. This elevated star formation is reflected in the higher specific star formation rates (sSFRs) of our XMPGs compared to other galaxies, consistent with previous studies (e.g., \citet{2013A&A...558A..18F, 2015ApJ...802...82F}). Their short stellar mass doubling timescales (likely much shorter than their ages) suggest that these systems are undergoing intense, likely unsustainable starburst episodes. This intense activity may be attributed to their early evolutionary stage and substantial gas reservoirs. To mitigate potential selection effects, we also perform the same comparison using a more complete, low-redshift subset of XMPGs and a matched control sample at $z<0.1$. This test reveals no difference in the observed trends.

To investigate whether the distinct SFMS of our XMPGs reflects properties of galaxies in the early universe, we compare the XMPG SFMS with a sample of 201 high‑redshift galaxies ($3.0 < z < 9.4$) observed by the James Webb Space Telescope (JWST) \citep{2023ApJS..269...33N, 2024A&A...684A..75C, 2025ApJ...978..136S}. The combined sample draws from three major studies: 64 galaxies with reliable SFRs and metallicities from \cite{2023ApJS..269...33N} (selected via strong $\mathrm{[O III]\lambda \lambda 4959,5007}$ and $\text{H}\beta$ emission from ERO, GLASS, and CEERS data at $z=3.8-8.9$); 62 galaxies from the deep GOODS-S tier of JADES \citep{2024A&A...684A..75C}, selected at $3<z<10$ with robust emission-line detections (e.g., $\mathrm{S/N \geq 5}$ for strong lines of $\text{H} \alpha$, $\text{H}\beta$ and $\mathrm{[O III]\lambda \lambda 4959,5007}$, and $\mathrm{S/N \geq 3}$ for weaker lines) to ensure reliably derivable gas-phase metallicity; and 75 galaxies from the GOODS-N/S fields and the PRIMAL survey \citep{2025ApJ...978..136S} at $4 < z < 10$, satisfying similar emission-line quality cuts. These JWST galaxies are all star-forming systems with prominent emission lines, from which objects with evident AGN activity have been excluded. They span stellar masses of $\log(M_{\star}/M_{\odot}) = 6.4 - 10.1$, SFRs of log(SFR/($M_{\odot}/yr$) = $-0.4 - 2.0$, and gas- phase metallicities of 12 + log(O/H) = $7.14-8.57$.

It is should be noted that the combined JWST sample is not a complete census. The selection is primarily based on the detectability of strong rest-optical emission lines, which may introduce inherent incompleteness and selection biases, especially at the low-mass end. However, the JWST sample provides a crucial high-redshift benchmark, the comparison is necessarily more informative for the higher-mass overlap region.
All metallicities of these high-redshift galaxies are calibrated to the $T_{\mathrm{e}}$ method. Stellar masses are primarily derived through SED fitting, while SFRs are mainly obtained from Balmer line fluxes, with a small fraction estimated from UV absolute magnitudes. To ensure a consistent comparison with our XMPGs, the SFRs are recalibrated accordingly. Additionally, stellar masses from the JWST samples are converted to a Chabrier initial mass function (IMF) using established conversions from \citet{2004MNRAS.351.1151B, 2024A&A...683A.184G, 2024MNRAS.532.2832L}, aligning with the IMF adopted in our DESI XMPG analysis.

These JWST galaxies are divided into three stellar mass intervals: $\log(M_{\star}/M_{\odot}) \leq 7.5 $, $7.5 < \log(M_{\star}/M_{\odot}) \leq 8.5$, and $\log(M_{\star}/M_{\odot}) > 8.5$, with the mean SFR and standard deviation computed within each bin (see Figure \ref{fig:XMPG_VS_Control_Sample}). The comparison reveals that high-redshift JWST galaxies follow an even flatter SFMS than local XMPGs. At $\mathrm{log}(M_{\star}/M_{\odot}) < 7.5$, the XMPG SFMS lies below the mean SFR of high-redshift JWST galaxies (the deviation of $\sim 2.5 \sigma$). This offset could reflect a genuine physical difference, but it may also be influenced by the incompleteness of the JWST samples at the low-mass end. When considering measurement uncertainties, the offset diminishes in the higher mass bins, where the XMPG SFMS converges with the high-redshift data. This convergence indicates that local XMPGs with ($\log(M_{\star}/M_{\odot}) > 7.5$) maintain SFRs and sSFRs comparable to those of high-redshift star-forming galaxies. 

\subsection{Fundamental Metallicity Relation} \label{subsec:FMR}
The mass-metallicity relation (MZR), which demonstrates the correlation between increasing galaxy metallicity and stellar mass, represents a fundamental empirical relation in galaxy evolution studies \citep{1979A&A....80..155L, 2004ApJ...613..898T}. Subsequent research expanded this two-dimensional relation by incorporating additional physical parameters. \cite{2008ApJ...672L.107E} discovered that for galaxies of fixed $\mathrm{M_{\star}}$, metallicity decreases with increasing sSFR and half-light radius, highlighting the importance of star formation and galaxy structure in regulating chemical enrichment.

Building on these insights, \cite{2010MNRAS.408.2115M} established the fundamental metallicity relation (FMR), which is a tight three-dimensional correlation linking gas-phase metallicity, $\mathrm{M_{\star}}$, and SFR. The MZR emerges as a projection of this more fundamental relation onto the $\mathrm{M_{\star}}$-metallicity plane. A defining characteristic of the FMR is its significantly reduced scatter compared to the MZR, as SFR captures the balance between metal production, gas inflow, and outflow processes \citep{2010MNRAS.408.2115M, 2013ApJ...765..140A}.  Crucially, the FMR demonstrates evolutionary invariance up to $z \sim 2.5$ \citep{2010MNRAS.408.2115M}, establishing it as a cornerstone for modeling chemical enrichment across most of cosmic history.  Subsequent studies have further constrained the FMR using large, homogeneous samples from the SDSS \citep{2013ApJ...765..140A, 2020MNRAS.491..944C}.

While the local FMR remains invariant up to $z \lesssim 2.5$, recent studies indicate significant evolution at higher redshifts \citep{2010MNRAS.408.2115M, 2024A&A...684A..75C, 2025ApJ...978..136S, 2025MNRAS.540.1800S}. Cosmological simulations support this picture, suggesting substantial FMR evolution in the high-redshift universe \citep{2024MNRAS.531.1398G, 2025MNRAS.536..119G}. The advent of JWST has enabled direct observational constraints on this evolution by providing metallicity measurements for galaxies out to $z \sim 10$, allowing direct comparison with FMR predictions.

\begin{figure*}[htbp!]
\centering
	\includegraphics[width=0.8\textwidth]{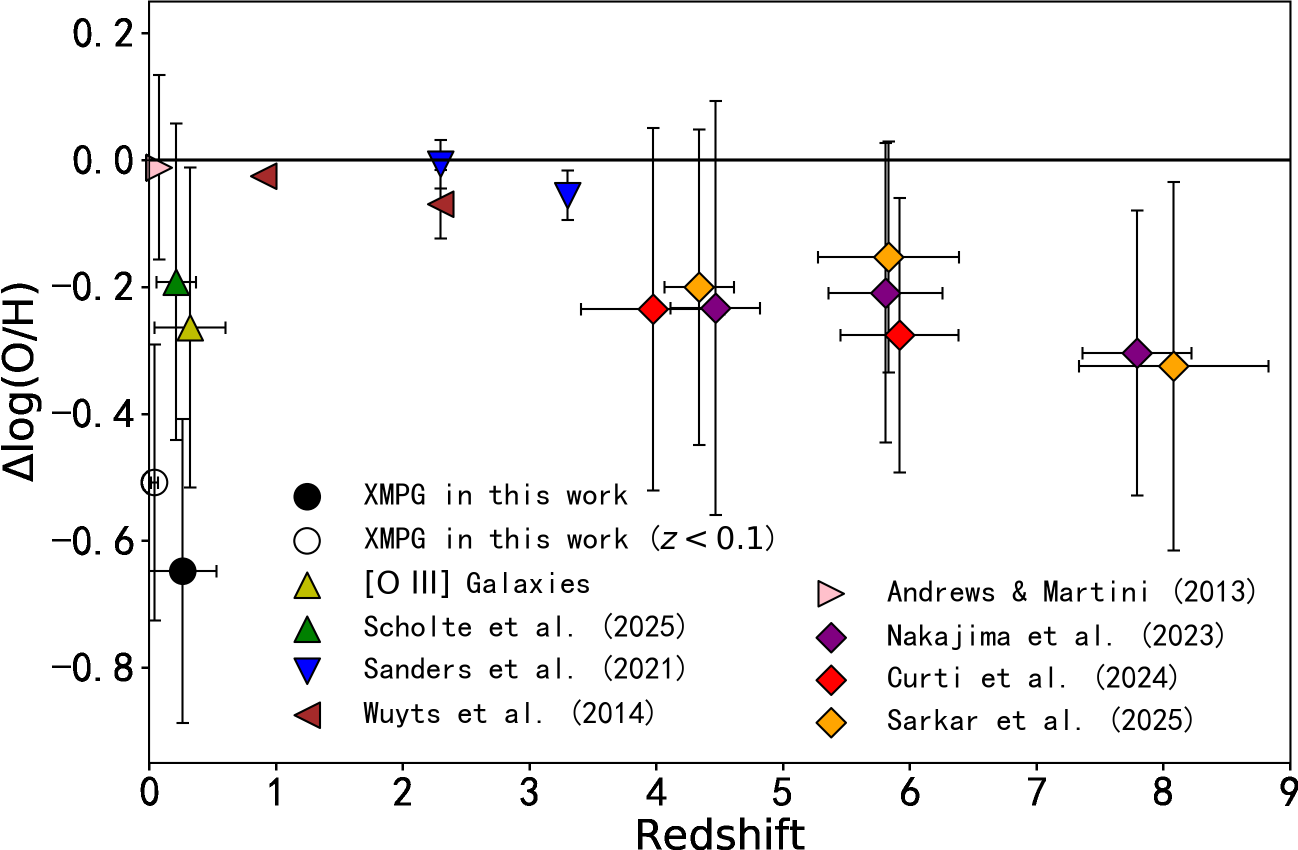}
    \caption{Comparison between observed metallicities and the predictions from the FMR of \cite{2020MNRAS.491..944C}. Markers indicate the mean redshift and metallicity offset $\Delta \mathrm{log(O/H)}$ (observed minus predicted), with error bars showing standard deviations. Data are shown for: the confirmed XMPGs from this work (black filled circle) and its $z<0.1$ subset (open circle); $T_e$-based metallicities from DESI EDR \citep[][green triangle]{2025MNRAS.540.1800S} and all 9,292 metallicities in this work (yellow triangle); stacked spectra from SDSS (\citet{2013ApJ...765..140A}; pink triangles), at $z\sim 0.9, 2.3$ (\citet{2014ApJ...789L..40W}; brown triangles), and at $z\sim 2.3,3.3$ (\citet{2021ApJ...914...19S}; blue triangles); and JWST high-redshift galaxies in redshift bins $z<5$, $5<z<7$, and $z>7$(diamonds: \citet{2023ApJS..269...33N} in purple, \citet{2024A&A...684A..75C} in red, \citet{2025ApJ...978..136S} in orange). For these JWST galaxies, the bins with fewer than 5 samples are excluded. The black horizontal line denotes zero offset.  \label{fig:metallicity_VS_FMR}}
\end{figure*}

Figure \ref{fig:metallicity_VS_FMR} compares the gas-phase metallicities of our XMPGs, other low- and intermediate-redshift galaxies, and JWST high-redshift galaxies against the FMR predictions as a function of redshift. All comparison metallicities are either derived via the $T_e$ method or are consistently calibrated to it. We adopt the FMR formulation derived by \cite{2020MNRAS.491..944C}, which is derived on a suitable low-redshift SDSS galaxy sample:
\begin{equation}
Z(M, \mathrm{SFR})=Z_{0}-\gamma / \beta \log \left(1+\left(M / M_{0}(\mathrm{SFR})\right)^{-\beta}\right),
\end{equation}
where $M_{0}(\mathrm{SFR})=10^{m_{0}} \mathrm{SFR}^{m_{1}}$, $Z_0=8.779 \pm 0.005$, $m_{0}=10.11 \pm 0.03$, $m_{1}=0.56 \pm 0.01$, $\gamma=0.31 \pm 0.01$, and $\beta= 2.1 \pm 0.4$. As shown in the figure, metallicities from low-redshift SDSS galaxies \cite{2013ApJ...765..140A} and galaxies at $z<1$ from \citet{2014ApJ...789L..40W} show excellent agreement with the local FMR prediction with negligible systematic offsets. Minor deviations between measured and predicted metallicities begin to emerge for galaxies at $2<z<3$ \citep{2014ApJ...789L..40W,2021ApJ...914...19S}, but these remain statistically insignificant. 

JWST observations reveal that galaxies at $3 < z < 10$ with stellar masses $\text{log}(M_{\star}/M_{\sun}) \sim 6.5-10$ are significantly more metal-poor than predicted by the local FMR. This metal deficiency exhibits a moderate redshift dependence: the median difference between galaxy metallicities and the FMR predicted values ($\Delta\text{log(O/H)}$) evolves from $-0.2 \pm 0.3$ dex at $z \sim 4$ to $-0.3 \pm 0.3$ at $z \sim 8$. Although \cite{2025ApJ...978..136S} reported that galaxies at $4 < z < 8$ show no such significant systematic if adopting different FMR form, galaxies at $z > 8$ still exhibit a significant decrease in metallicity.

In Figure \ref{fig:metallicity_VS_FMR}, JWST high-redshift galaxies generally exhibit metallicities lower than FMR predictions by approximately 0.2 dex. At $z>7$, the mean offset in metallicity exceeds the 1$\sigma$ scatter, indicating significant FMR evolution with redshift. We further compare the metallicities of our [O III]$\lambda 4363$-detected galaxies (including our XMPGs) with the predictions of the FMR. Their metallicities are systematically lower than the local FMR, which aligns with the offsets observed in JWST high-redshift galaxies. Our XMPGs with a mean redshift of 0.26 exhibits a substantial offset from the local FMR, characterized by a mean deviation of $-0.65 \pm 0.24$ dex. This offset is significantly larger than that observed in high-redshift JWST galaxies. Even when restricting the analysis to the more complete subset of XMPGs with $z<0.1$, a considerable offset of $-0.50 \pm 0.22$ persists, still markedly exceeding the JWST values. Notably, the metallicity offsets of JWST high-redshift galaxies show a decreasing trend with increasing redshift. Extrapolating this trend suggests that our local XMPGs exhibit deviations comparable to galaxies at redshifts beyond the current reach of JWST. This positions local XMPGs as potential analogs to extremely primitive galactic systems in the early Universe. 

The extremely low metallicities of XMPGs are likely attributable to several concurrent physical mechanisms. These can operate in combination or independently to arrest chemical enrichment.
\begin{enumerate}
\item \textbf{A primordial origin:} XMPGs might be cosmic "living fossils," having condensed from gas clouds that experienced few prior episodes of star formation. Consequently, their interstellar medium has undergone minimal chemical enrichment \citep{2024arXiv241204541N}.
\item \textbf{Dilution by pristine gas accretion:} The distinct irregular "cometary" morphology of some XMPGs signals recent accretion of metal-poor gas from cosmic web \citep{2021ApJ...918...54I}. This infall dilutes the metallicity of the interstellar medium \citep{2015ApJ...802...82F,2024arXiv241204541N}. The associated compression of the galactic gas can also trigger the intense starbursts commonly observed in these systems.
\item \textbf{Inefficient enrichment from galactic outflows:} As low-mass and gas-rich dwarfs, most XMPGs possess shallow gravitational potentials \citep{2015Natur.523..169E,2022ApJ...929..134X}. This makes them highly susceptible to stellar feedback (winds and supernovae), which can efficiently eject metal-enriched gas into the circumgalactic medium. This process continuously removes newly synthesized metals, preventing the enrichment of the remaining gas reservoir. 
\end{enumerate}

\section{Conclusions} \label{sec:summary}
This study presents the largest catalog of XMPGs to date, identified and analyzed using DESI DR1 data. The sample was constructed by systematically selecting 9,372 galaxies with secure detections of the temperature-sensitive auroral line $\mathrm{[O\ III]} \lambda 4363$, reliable stellar masses, and no AGN contamination. Gas-phase metallicities were derived for 9,292 sources using the direct $T_e$-method, from which 1,475 low-metallicity candidates were identified. After visual inspection to remove 52 contaminants, 656 galaxies were classified as confirmed XMPGs (where the upper metallicity uncertainty bound remains below 12+log(O/H) $=$ 7.69) threshold, with 767 sources designated as high-quality candidates. The final sample, after accounting for previous discoveries, yields 1,221 novel XMPGs (551 confirmed and 670 candidates), representing a substantial expansion of the known population.

The most metal-poor galaxy in our sample (DESI J093402.37+551423.2) corresponds to the southeast component of the well-known XMPG I Zw 18. With 12+log(O/H) $= 6.99 \pm 0.07$, it approaches the lower metallicity limit for low-redshift galaxies ($\sim 2 \%$ of solar). This low-mass starburst system exhibits strong emission lines and a young stellar continuum. Together with its northwest companion located 0.47 kpc away, these two components share similar properties, indicating a possible common metal-poor gas origin.

Our analysis of the SFMS and FMR reveals that these local XMPGs are not merely metal-poor but exhibit scaling relations characteristic of galaxies in the early universe. Their SFMS shows enhanced normalization and a shallower slope compared to normal star-forming galaxies, indicating heightened star formation efficiency with a weaker mass dependence. For galaxies with log$(M_\star)>7.5$, the XMPG SFMS aligns with that of high-redshift galaxies observed by JWST. Furthermore, both the metallicities of the local XMPGs and the JWST high-redshift galaxies are systematically lower than predictions from local FMRs, with the deviation being more pronounced for the XMPGs. This consistent offset suggests that local XMPGs retain physical properties, likely driven by efficient gas accretion, akin to primitive galaxies at high redshift.

\begin{acknowledgments}
The authors acknowledge the supports from the National Natural Science Foundation of China (NSFC; grant Nos. 12120101003, 12373010, 12173051, 12233008, and 12503019), and China Manned Space Project (No. CMS-CSST-2025-A06). The authors also acknowledge the National Key R\&D Program of China (grant Nos. 2022YFA1602902, 2023YFA1607804, 2023YFA1607800 and 2023YFA1608100), Strategic Priority Research Program of the Chinese Academy of Sciences with Grant Nos. XDB0550100 and XDB0550000 and Programs of the National Astronomical Observatories, the Chinese Academy of Sciences (Grant Nos. E5ZQ7801, E5ZB7801, and E4TG2001).

M.S. acknowledges support by the State Research Agency of the Spanish Ministry of Science and Innovation under the grants 'Galaxy Evolution with Artificial Intelligence' (PGC2018-100852-A-I00) and 'BASALT' (PID2021-126838NB-I00) and the Polish National Agency for Academic Exchange (Bekker grant BPN/BEK/2021/1/00298/DEC/1). This work was partially supported by the European Union's Horizon 2020 Research and Innovation program under the Maria Sklodowska-Curie grant agreement (No. 754510).

This material is based upon work supported by the U.S. Department of Energy (DOE), Office of Science, Office of High-Energy Physics, under Contract No. DE–AC02–05CH11231, and by the National Energy Research Scientific Computing Center, a DOE Office of Science User Facility under the same contract. Additional support for DESI was provided by the U.S. National Science Foundation (NSF), Division of Astronomical Sciences under Contract No. AST-0950945 to the NSF’s National Optical-Infrared Astronomy Research Laboratory; the Science and Technology Facilities Council of the United Kingdom; the Gordon and Betty Moore Foundation; the Heising-Simons Foundation; the French Alternative Energies and Atomic Energy Commission (CEA); the National Council of Humanities, Science and Technology of Mexico (CONAHCYT); the Ministry of Science, Innovation and Universities of Spain (MICIU/AEI/10.13039/501100011033), and by the DESI Member Institutions: \url{https://www.desi.lbl.gov/collaborating-institutions}. Any opinions, findings, and conclusions or recommendations expressed in this material are those of the author(s) and do not necessarily reflect the views of the U. S. National Science Foundation, the U. S. Department of Energy, or any of the listed funding agencies.

The authors are honored to be permitted to conduct scientific research on I'oligam Du'ag (Kitt Peak), a mountain with particular significance to the Tohono O’odham Nation.
\end{acknowledgments}

\appendix

\section{Appendix}
In our work, 656 confirmed XMPGs and 767 XMPG candidates are identified. The properties of a representative subset of these galaxies, including redshift, metallicity, stellar mass, and star formation rates, etc., are provided in Appendix Tables \ref{tab:conformed_XMPG} and \ref{tab:XMPG_candidate}.

The data points of the figures and the tabular data presented in this paper are available for download at the Zenodo DOI \url{https://doi.org/10.5281/zenodo.18309897}.

\begin{splitdeluxetable*}{crrcccccBcccccc}
\tabletypesize{\scriptsize}
\tablewidth{0pt} 
\tablecaption{Properties of Confirmed XMPGs from DESI DR1 \label{tab:conformed_XMPG}}
\tablehead{
\colhead{No} & \colhead{Name}& \colhead{TARGETID} & \colhead{R.A.} & \colhead{Flux} & \colhead{Redshift} & \colhead{$r$} & \colhead{E(B-V)} & \colhead{EW([O III])} & \colhead{$N_e$} & \colhead{$T_e(\mathrm{[O\ III]})$} & \colhead{12+log(O/H)} & \colhead{log($M_{\star}/M_{\odot}$)} & \colhead{log(SFR/($M_{\odot}/yr$))}
\\
 & & & \colhead{(deg)} & \colhead{(deg)} & & \colhead{(mag)} & \colhead{(mag)} & \colhead{($\mathrm{\AA}$)} & \colhead{(cm$^{-3}$)} & \colhead{(K)} & & & 
}
\colnumbers
\startdata
    1 &  DESIJ093402.37+551423.2 &  39633324993414901 & 143.50987 &  55.23980 & 0.0026 & 17.53 &            0.00 & 262.09 & 100 & 22807$\pm$1868 & 6.99$\pm$0.07 & 5.44$\pm$0.15 & -2.03$\pm$0.13 \\  
    2 &  DESIJ164114.56+470721.3 &  39633200686828028 & 250.31067 &  47.12260 & 0.1507 & 22.03 & 0.08 $\pm$ 0.10 & 988.55 & 100 & 26857$\pm$2538 & 7.02$\pm$0.06 & 7.03$\pm$0.11 & 0.06$\pm$0.13 \\  
    3 &  DESIJ001859.31+234540.3 &  39628343171090153 &   4.74714 &  23.76121 & 0.0154 & 19.07 &            0.00 & 122.16 & 100 & 23693$\pm$3072 & 7.06$\pm$0.14 & 6.78$\pm$0.27 & -1.35$\pm$0.13 \\  
    4 &  DESIJ141046.05+002629.6 &  39627800373628335 & 212.69188 &   0.44156 & 0.0174 & 19.34 &            0.00 & 55.95 & 100 & 28683$\pm$7892 & 7.06$\pm$0.17 & 7.22$\pm$0.18 & -1.79$\pm$0.13 \\  
    5 &  DESIJ222003.53+282422.3 &  39628451342195230 & 335.01469 &  28.40622 & 0.0120 & 19.07 & 0.01 $\pm$ 0.04 & 145.48 & 100 & 24448$\pm$5868 & 7.07$\pm$0.16 & 6.59$\pm$0.26 & -1.61$\pm$0.13 \\ 
\enddata
\tablecomments{(1) Serial Number; (2) Galaxy name in DESI; (3) TARGETID of the galaxy in DESI; (4) Right Ascension (RA); (5) Declination (Dec); (6) Redshift; (7) $r$-band magnitude; (8) gas-phase extinction in $E(B-V)$; (9) Equivalent width of [O III]$\lambda 5007$; (10) Electron density; (11) [O III] electron temperature; (12) Metallicity; (13) Stellar mass; (14) Star formation rate.}
\end{splitdeluxetable*}

\begin{splitdeluxetable*}{crrcccccBcccccc}
\tabletypesize{\scriptsize}
\tablewidth{0pt} 
\tablecaption{Properties of XMPG Candidates from DESI DR1 \label{tab:XMPG_candidate}}
\tablehead{
\colhead{No} & \colhead{Name}& \colhead{TARGETID} & \colhead{R.A.} & \colhead{Flux} & \colhead{Redshift} & \colhead{$r$} & \colhead{E(B-V)} & \colhead{EW([O III])} & \colhead{$N_e$} & \colhead{$T_e(\mathrm{[O\ III]})$} & \colhead{12+log(O/H)} & \colhead{log($M_{\star}/M_{\odot}$)} & \colhead{log(SFR/($M_{\odot}/yr$))}
\\
 & & & \colhead{(deg)} & \colhead{(deg)} & & \colhead{(mag)} & \colhead{(mag)} & \colhead{($\mathrm{\AA}$)} & \colhead{(cm$^{-3}$)} & \colhead{(K)} & & & 
}
\colnumbers
\startdata
    1 &  DESIJ032851.16-045421.2 &  39627664897607093 &  52.21316 &  -4.90590 & 0.0818 & 19.27 & 0.19 $\pm$ 0.42 & 103.07 & 100 & 22442$\pm$5478 & 7.49$\pm$0.21 & 8.54$\pm$0.21 & 0.03$\pm$0.13 \\  
    2 &  DESIJ113601.71+382543.9 &  39633048681054957 & 174.00711 &  38.42888 & 0.8892 & 22.47 &            0.00 & 191.92 & 100 & 22532$\pm$5193 & 7.50$\pm$0.19 & 9.29$\pm$0.28 & 0.80$\pm$0.14 \\  
    3 &  DESIJ223659.88+010906.8 &  39627820615344672 & 339.24950 &   1.15190 & 0.3201 & 22.40 & 0.05 $\pm$ 0.08 & 589.50 & 100 & 21081$\pm$4702 & 7.51$\pm$0.20 & 9.00$\pm$0.14 & -0.01$\pm$0.13 \\  
    4 &  DESIJ144642.19+052129.7 &  39627915180115324 & 221.67580 &   5.35826 & 0.8774 & 23.75 & 0.70 $\pm$ 0.30 & 782.80 & 100 & 23005$\pm$5462 & 7.51$\pm$0.19 & 9.60$\pm$0.29 & 1.62$\pm$0.13 \\  
    5 &  DESIJ131216.32+021742.5 &  39627842408941092 & 198.06800 &   2.29516 & 0.0297 & 17.76 &            0.00 & 174.29 & 100 & 20890$\pm$4249 & 7.51$\pm$0.19 & 7.80$\pm$0.18 & -0.51$\pm$0.13 \\
\enddata
\tablecomments{The content of each column is the same as that in Table \ref{tab:conformed_XMPG}.}
\end{splitdeluxetable*}


\bibliography{DESI_DR1_XMPG}{}
\bibliographystyle{aasjournalv7}



\end{document}